\documentclass[dvips]{acta}
\usepackage{supertabular,lscape,epsfig}
\usepackage{amssymb}
\usepackage{amsmath}
\usepackage[T1]{fontenc}

\SetPages{0}{0}

\SetVol{76}{2026}

\usepackage{lmodern}

\newcommand{\WT}{W_{\rm t}}
\newcommand{\PHX}{B_{\rm p}}
\newcommand{\PHS}{B_{\rm surf}}
\newcommand{\PHB}{B_{\rm bot}}
\newcommand{\RES}{\Re(\sigma)}
\newcommand{\IMS}{\Im(\sigma)}
\newcommand{\PTOT}{\mathcal{P}}
\newcommand{\EKIN}{\mathcal{E}_{\rm kin}}
\newcommand{\PEX}{\mathcal{P}_{\rm ex}}
\newcommand{\PRSP}{\mathcal{P}_{\rm rsp}}
\newcommand{\QEX}{\mathcal{Q}_{\rm ex}}
\newcommand{\QTOT}{\mathcal{Q}_{\rm tot}}
\newcommand{\PDB}{\Delta B}
\newcommand{\NMC}{N_{\rm c}}
\newcommand{\NMH}{N_{\rm ha}}
\newcommand{\NMS}{N_{\rm sg}}
\newcommand{\NMI}{N_{\rm in}}
\newcommand{\NEF}{N_{\rm eff}}
\newcommand{\CHIE}{\chi_{\rm eff}}
\newcommand{\LTEF}{\log(T_{\rm eff})}

\newcommand{\mFig}[1]{Fig.~\ref{fig:#1}}
\newcommand{\mEq}[1]{Eq.~(\ref{eq:#1})}

\begin{document}

\begin{Titlepage}

\Title{Thermodynamic exchange and mode stability in strongly nonadiabatic radial pulsations}

\Author{Zalewski, J.}
{Independent researcher \\
e-mail: jan.zalewski.a2@gmail.com}

\end{Titlepage}

\Abstract{
We apply the quadratic balance relation for linear nonadiabatic pulsations to radial modes of a sequence of post-AGB envelope models covering $3.52\leq\log(T_{\rm eff})\leq4.6$. The relation separates the thermodynamic exchange power $\mathcal{P}_{\rm ex}$, response power $\mathcal{P}_{\rm rsp}$, and surface contribution $\Delta B$, and introduces an effective norm through $\mathcal{P}_{\rm ex}+\Delta B=2\gamma\chi_{\rm eff}\mathcal{E}_{\rm kin}$, where $\gamma$ is the mode growth rate. For nearly adiabatic modes $\chi_{\rm eff}$ is close to unity, whereas in strongly nonadiabatic pulsations the compression and horizontal area deformation contributions can make it change magnitude or even reverse its sign. In the cooler models, a broad family of overtone modes has $\chi_{\rm eff}<0$ and is damped despite positive thermodynamic exchange power. For these modes the response power supplies the opposing damping. Conversely, two low frequency modes in the hotter models are excited over a range of effective temperatures, even though their thermodynamic exchange power is negative. Their instability is produced by the positive response power. Strange modes generally have $\chi_{\rm eff}>0$, apart from a narrow boundary-sensitive transition during which a cold strange mode transforms into an ordinary p-mode. The excited low frequency solution with $\chi_{\rm eff}<0$ persists with a finer numerical mesh and such modes are found under the alternative outer boundary formulations examined, although their frequencies are boundary condition dependent. Thus, for strongly nonadiabatic pulsations, the sign of the thermodynamic exchange power alone does not determine mode stability.
}
{stars: AGB and post-AGB, stars: oscillations, stars: interiors, instabilities, methods: analytical, methods: numerical}

\section{Introduction}
The relation between pulsation mode thermodynamic work and the rate of its kinetic energy change (Cox 1980, Unno et al. 1989) is often used to identify the stellar regions responsible for excitation and to locate instability domains on the HR diagram. For example, Dziembowski (1994) examined instability domains of hot B stars, while Saio et al. (1998) used work curves to identify driving and damping regions of strange modes.

The customary interpretation of the cycle averaged thermodynamic work in terms of the pulsation growth rate assumes that the mode amplitude changes little during an oscillation period. Gautschy \& Glatzel (1990) noted that this assumption fails for strongly nonadiabatic modes, whose growth or damping times may be comparable to the oscillation period. Buchler \& Regev (1982) considered an extension based on a multiple-time formalism, while Glatzel (1994) derived an ensemble-averaged balance relation for strongly nonadiabatic pulsations. 

Strange mode instabilities that could not be attributed to the classical $\kappa$ mechanism were identified in pulsations in luminous stellar envelopes (Gautschy \& Glatzel 1990, Glatzel 1994). Gautschy \& Glatzel associated such instabilities with mode coupling, while Glatzel interpreted them in terms of a phase lag between pressure and density perturbations. Saio et al. (1998) also found that, in extremely nonadiabatic strange modes, a radiation force perturbation out of phase with the density perturbation can produce overstability. In full nonadiabatic calculations, Sonoi \& Shibahashi (2014) found unstable strange modes without adiabatic counterparts whose driving regions were not confined to those satisfying the usual criterion for the $\kappa$ mechanism. They also identified a mode in which $\kappa$ mechanism driving and strange mode instability acted together. These results show that mode instability need not be attributable to the $\kappa$ mechanism alone. They do not, however, address the distinct question considered here: whether the global thermodynamic exchange power and the mode growth rate can have opposite signs once the remaining contributions to the pulsation balance are separated.

The present analysis does not invoke any of the above mentioned forms of approximations or assume a particular excitation mechanism. Instead, the thermodynamic exchange and response contributions are determined from an exact quadratic balance relation evaluated for solutions of the full linear nonadiabatic equations. In our previous work (Zalewski 2026c, 2026d) we have shown that the balance relation between the mode power and its kinetic energy for nonadiabatic pulsation includes several competing terms in addition to the classical thermodynamic work term and the inertia term. From these analyses it follows that the rate of change of mode kinetic energy is proportional to the total power, while the total power is not necessarily identical to the thermodynamic one. The balance relation for nonadiabatic, nonradial pulsation established in Zalewski (2026d) relates the exchange power, the response power and surface terms to the kinetic energy rate. These terms need to be taken into account in order to obtain the mode's excitation rate.

Using the formalism of Zalewski (2026d) it is possible to relate mode exchange (thermodynamic) power with rate of change of mode kinetic energy by introducing an effective norm. In what follows it will be shown that the effective norm factor, while close to unity for near-adiabatic pulsation, may deviate substantially and even become negative for strong nonadiabaticity. We identify the factors that contribute to the effective norm and the role of compression as well as horizontal area deformation terms for various types of radial nonadiabatic modes in the envelopes of post-AGB stars.

We find modes that are damped despite positive thermodynamic exchange power, as well as modes that are excited despite negative thermodynamic exchange power. These aspects will be presented and discussed in the subsequent sections following introduction of notation used in the present paper.

\section{Effective norm and exchange power balance}
In Zalewski (2026d) a balance relation between total power $\PTOT$ and the rate of change of kinetic energy $\EKIN$ was derived for linear nonadiabatic pulsations, both for the radial as well as nonradial cases. This relation is given by
\begin{equation}
\PTOT+\PDB=2\RES \EKIN,
\label{eq:PTOTEK}
\end{equation}
where
\[
\PTOT = \PEX + \PRSP,
\]
is the sum of the exchange power and the response power, and
\[
\PDB=\PHB-\PHS,
\]
is the difference of the surface terms, each of which is given by
\[
B=\PHX+B_{\rm g},
\]
where
\[
\begin{aligned}
\PHX &=C\, \Re\left(p\,\overline{\sigma d}\right), \\
B_{\rm g}&= \Im(\sigma)\Im(\mathcal{G}).
\end{aligned}
\]
The $p$ is the Lagrangian pressure perturbation, $d$ is the radial displacement perturbation, $h$ is the horizontal displacement perturbation, $s$ is entropy perturbation and $\sigma$ is the nondimensional complex pulsation frequency related to pulsation frequency $\omega$ as $\omega=\sqrt{4\pi G\langle\rho\rangle}\,\sigma$, and the time dependence of perturbed quantities is $\exp(\omega t)$. The other terms used are given in Zalewski (2026d).

The power terms are given by
\[
\begin{aligned}
\PEX&=-\WT,\\
\PRSP&=\mathcal{P}_{\rm c}+\mathcal{P}_{\rm ha}+\mathcal{P}_{\rm sg},
\end{aligned}
\]
where
\[
 \WT=\int_{x_b}^{x_s} C(x) A_7 \Re(p \, \overline{\sigma s}) \, dx,
\]
is the thermodynamic work integral of the mode.

The response power ($\PRSP$) is a sum of the compression, the horizontal area deformation and the gravitational-stratification response power terms respectively. These terms are given by
\[
\begin{aligned}
	\mathcal{P}_{\rm c} &=\Re(\sigma) \NMC, \\
	\mathcal{P}_{\rm ha}&=\Re(\sigma) \NMH, \\
	\mathcal{P}_{\rm sg}&=\Re(\sigma) \NMS.
\end{aligned}
\] 
As it is seen the response power is proportional to the $\gamma=\Re(\sigma)$ and depends on the norm terms defined as
\[
\begin{aligned}
	\NMC &= -\int_{x_b}^{x_s} C\, A_4 |p|^2 \, dx, \\
	\NMH &= 2\int_{x_b}^{x_s} C\, A_3\Re\left(d\,\overline{\epsilon_A}\right)\, dx, \\
    \NMS &= -\int_{x_b}^{x_s} C\, A_3\left[\Re{\left(w_1\,\overline{d}+\Lambda w \,\overline{h}\right)}+A_5|d|^2\right]\, dx,
\end{aligned}
\]
where $\epsilon_A=2d-\Lambda h$.

The kinetic energy $\EKIN$ of the mode may be expressed in terms of the norm $\NMI$ representing the mode inertia as
\[
\EKIN=-\frac{\NMI}{2},
\]
where
\[
\NMI=-\int_{x_b}^{x_s} C\,A_3A_2|\sigma|^2\left(|d|^2+\Lambda|h|^2\right)\,dx,
\]
$\Lambda=\ell(\ell+1)$ and $x_b$ and $x_s$ denote the location of the inner and outer boundary of the pulsation region, with the independent variable $x=\ln(r/R_{\odot})$.

Combining the equations it is obtained that
\[
\PEX+\Delta B=-\gamma \left(\NMI+\NMC+\NMH+\NMS\right).
\]
This equation relates the power terms - the thermodynamic $\PEX$ and the difference of surface terms $\Delta B$ with the norm terms and the excitation rate. This equation may be written using the effective norm $\NEF=\NMI+\NMC+\NMH+\NMS$ as
\[
\PEX+\Delta B=-\gamma \NEF.
\]

We may introduce a dimensionless effective-norm factor $\CHIE$ defined as
\[
\CHIE = \frac{\NEF}{\NMI}=1+\frac{\NMC+\NMH+\NMS}{\NMI},
\]
using which it is possible to write the balance relation as
\begin{equation}
\begin{aligned}
\QEX\equiv\PEX+\PDB&=-\gamma\NMI\CHIE \\
&=2\gamma \EKIN \CHIE.
\end{aligned}
\label{eq:PEXEK}
\end{equation}

Since all the norm terms scale quadratically with eigenfunction
amplitude, $\CHIE$ and the individual ratios
$\NMC/\NMI$, $\NMH/\NMI$, and $\NMS/\NMI$ are independent
of eigenfunction normalization. The component ratios indicate how
the respective response terms modify the effective norm relative to
the inertial contribution. Because the signs of $\NMH$ and $\NMS$ depend of the mode (see Zalewski 2026d) they may lead a sign change of $\CHIE$.

Since $\EKIN>0$, Eq.~(\ref{eq:PEXEK}) shows that for
$\CHIE>0$ the exchange-plus-boundary power and the growth rate
$\gamma$ have the same sign, whereas for $\CHIE<0$ their signs
are opposite. If $\CHIE=0$, the balance relation requires
$\PEX+\PDB=0$ and becomes degenerate as an equation for $\gamma$.

While \mEq{PTOTEK} relates total power (and surface terms) to the kinetic energy of the mode, the \mEq{PEXEK} relates the thermodynamic power plus surface terms to the kinetic energy using $\CHIE$.

The classical form of the relation is recovered formally by neglecting the response-term contributions $\NMC$, $\NMH$ and $\NMS$, in which case $\CHIE=1$. And if the boundary contributions are negligible, the relation reduces to $\PEX=2\gamma\EKIN$.

For radial pulsation the $\CHIE$ is obtained by setting the term $\NMS=0$ and $\NMH=N_{\rm str}$ (see Zalewski 2026d). 

As it is seen from \mEq{PEXEK} the relation between the thermodynamic power ($\PEX$) and surface terms ($\PDB$) and the kinetic energy of the mode for nonadiabatic pulsation is similar to the classical relation provided the factor $\CHIE$ is retained. This factor depends on the mode structure and since the norm terms entering $\CHIE$ need not be sign-definite the effective-norm factor can become negative thus affecting the relation between the exchange power of the mode (including surface terms) and mode growth rate.

In the next section we will analyze the behavior of $\CHIE$ for ordinary radial p-modes as well as strange modes in the envelopes of post-AGB stars.

\section{Application to envelope modes}

For the purpose of examination of the $\CHIE$ behavior across a range of effective temperatures for models of post-AGB envelopes we have adopted a $M=0.69 M_{\odot}$, $L=10^4 L_{\odot}$ parameters and used the pulsation code described in Zalewski (2026b) using the continuous renormalization together with tracking transformation to compute the eigenfrequencies and eigenmodes. We shall analyze the behavior of $\CHIE$ for radial modes ($\ell=0$). Therefore the boundary conditions for all the models were assumed to be $(3,4)\text{--}(1,3)$ (see Zalewski 2026a, 2026d). We have focused on radial modes because their spectra are more stable with respect to the selection of the form of inner boundary conditions and their placement in the envelope. The spectrum and eigenmodes computations covered a range of effective temperatures $3.52 \le \LTEF\le 4.6$. Mixing length convection was assumed in the envelope model, but no provision for dynamic effects of convection, particularly on the lower temperature modes was made.

\begin{figure}[htb]
	\includegraphics{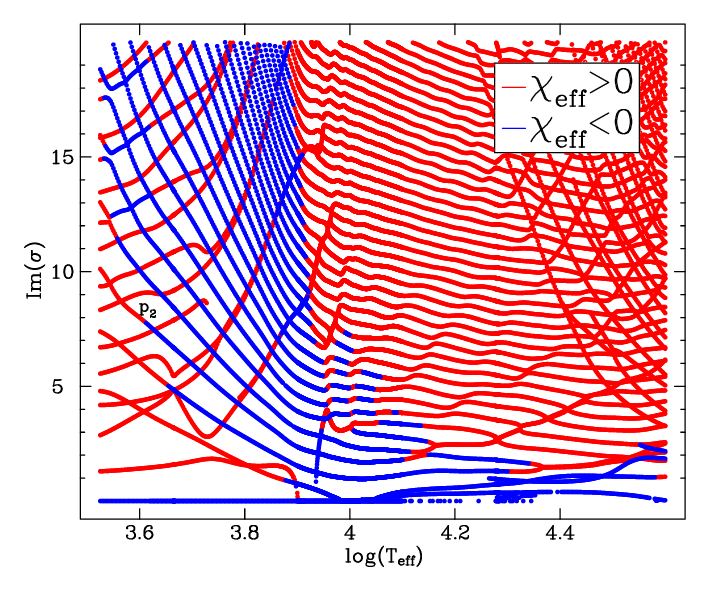}
	\FigCap{The plot of $\IMS$ versus $\LTEF$ for radial modes in a model $M=0.69 M_{\odot}$, $L=10^4 L_{\odot}$. Several families of modes are visible, grouping into two ranges - at effective temperatures lower than $4.0$ and at temperatures higher than $4.2$. The label $p_{2}$ denotes the second overtone p-mode discussed in the text. }
	\label{fig:Fig1}
\end{figure} 

The plot in \mFig{Fig1} shows several families of modes. Ordinary p-modes span the entire range of effective temperatures, with their frequencies decreasing with the increase of temperature. At $\LTEF<4.0$ strange mode branches whose $\Im(\sigma)$ increases with effective temperature are present. At temperatures $\LTEF>4.2$  a family of strongly damped modes appears with frequencies decreasing with the increase of temperature. These modes have large amplitudes in the opacity Z-bump region. In addition to these types of modes there are also two pairs of strange modes at higher effective temperatures at frequencies of low overtone p-modes, and thermal-oscillatory modes of frequencies below the fundamental mode. In addition there are thermal modes which however will not be followed further in this paper. 

The overall structure of the spectrum may be compared with the modal diagram obtained by Gautschy (1993) in his survey of radial pulsations in post-AGB envelope models over the range $3.8\lesssim \LTEF\lesssim 4.9$, which likewise contain several mode families whose frequencies either increase or decrease with effective temperatures. Gautschy (1993) also examined the influence of outer boundary conditions on the modal spectra of luminous stellar envelopes. The reflective outer boundary condition used in that work yields a strange-mode spectrum broadly comparable to that obtained with the $(3,4)$ outer boundary condition adopted here (Zalewski 2026a).

The frequency points shown in \mFig{Fig1} are color-coded as: red dots are the modes for which $\CHIE>0$ while blue ones are those for which $\CHIE<0$. The sign reversal of $\CHIE$ in \mEq{PEXEK} does not imply a sign change in the mode excitation rate, as this is determined by the boundary-inclusive total power $\mathcal{P}+\Delta B$ and mode kinetic energy $\EKIN>0$.

The compression term $\NMC$ is large in the red-marked areas, while the horizontal area deformation term $\NMH$ is dominating in the blue colored ones. 

From \mFig{Fig1} it may be seen that the region where $\CHIE<0$ forms an organized manifold which includes ordinary modes of high frequency at low effective temperatures, and it shifts to low frequency modes near $\LTEF=4$ and subsequently continues for lowest frequency modes to higher temperatures. Also the thermal modes are characterized by negative $\CHIE$. The low temperature strange modes ($\LTEF<4$), but also the high temperature ones ($4.2 \le \LTEF\le 4.6$) exhibit positive $\CHIE$. We shall not analyze in detail the families of damped modes for $\LTEF>4.2$ visible in \mFig{Fig1} as this would require extending the survey to much higher temperatures than the ones considered here. In the range of effective temperatures included in \mFig{Fig1} these modes have $\CHIE>0$.

\subsection{Sign reversal of $\CHIE$ along the second-overtone p-mode sequence at low $T_{\rm eff}$}

In order to analyze the change of the sign of $\CHIE$ it is convenient to introduce normalized cumulative forms as
\[
n_{\rm j}(t)=\frac{N_{\rm j}(t)}{N_{\rm in}(t_{\rm b})},\quad\quad j\in \{{\rm in},{\rm c},{\rm ha}\},
\]
where $t=\log(T)$ and $t_b$ is the temperature at the inner boundary of the pulsation region, $N_{\rm j}(t)$ are accumulated from the surface inwards and $\CHIE(t)$ is a cumulative diagnostic whole value $\CHIE(t_b)$ equals the global $\CHIE$. Using these it is possible to write
\[
\CHIE(t)=n_{\rm in}+n_{\rm c}+n_{\rm ha}.
\]
With the adopted normalization the $0\le n_{\rm in}\le 1$, and $n_{\rm in}(t_{\rm b})=1$, while the $n_{\rm c}$ and $n_{\rm ha}$ are the cumulative forms of pressure and horizontal area deformation terms normalized by the total mode inertia. For radial modes these two terms are of opposite sign.

\begin{figure}[htb]
	\includegraphics{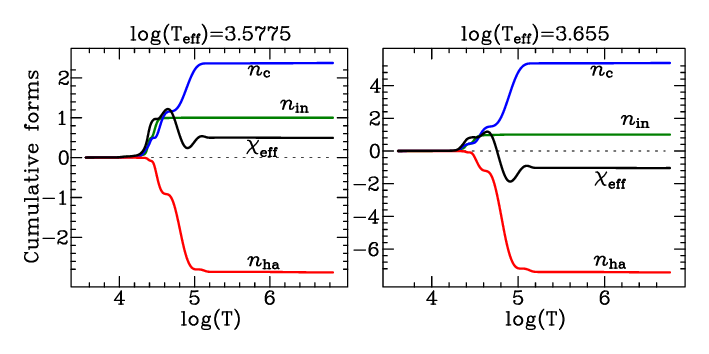}
	\FigCap{The normalized cumulative forms dependence on model temperature for two models with $\LTEF=3.5775$ and $3.655$ in the left and right panels respectively. It is seen that for the hotter model the horizontal area deformation term dominates combined inertia and compression terms making the $\CHIE<0$.}
	\label{fig:Fig2}
\end{figure}

For the second overtone p-mode ($p_{2}$) the sign of $\CHIE$ changes near $\LTEF\sim 3.6$ hence the normalized cumulative forms are shown in \mFig{Fig2} for two effective temperatures. For $\LTEF=3.5775$ the value of $\CHIE=0.49$, while for $\LTEF=3.655$ the $\CHIE=-1.04$.

With increasing effective temperature, both compressional and area-deformation responses strengthen, but the latter grows sufficiently rapidly to overcome the combined inertial and compressional contributions within the layers around $\log(T)\simeq 4.3 - 5.1$, causing the effective norm to change sign.

\subsection{Behavior of $\CHIE$ for strange modes}

The mode $p_{2}$ can be followed across the whole effective temperature range considered here, from $\LTEF=3.52$ to $4.6$, and its frequency evolution along the sequence is visible in \mFig{Fig1}. At the lowest temperatures the compression contribution dominates the $\CHIE$ . With increasing temperature, the horizontal area deformation term becomes important, until the compression term starts to dominate again at $\LTEF\approx 4.365$. 

Near this temperature, $p_{2}$ encounters the $p_{1}$ mode. Their oscillation frequencies, given by the imaginary parts of $\sigma$, become nearly equal, whereas their growth rates, given by the real parts, separate and attain large values of opposite sign. For $\LTEF\gtrsim 4.365$ the two modes can be followed as an excited and damped high temperature pair of strange modes. High temperature strange modes in post-AGB envelope sequences were previously discussed by Gautschy (1993), while related strange mode spectra in massive star models were studied by Glatzel \& Kiriakidis (1993) and Saio (2011).

\begin{figure}[htb]
	\includegraphics{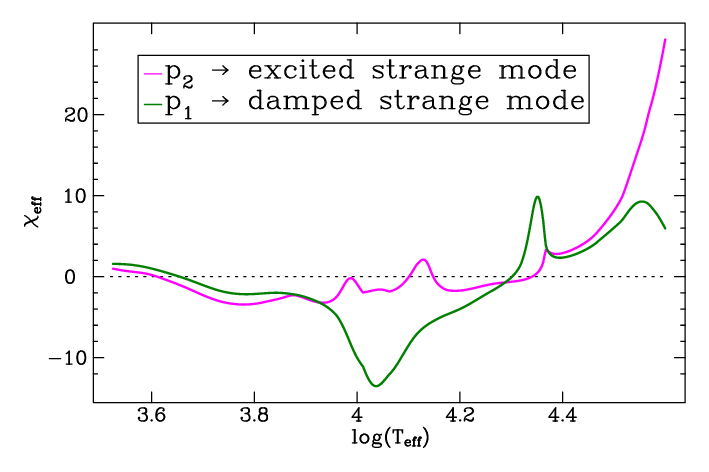}
	\FigCap{The dependence of $\CHIE$ on $\LTEF$ for the $p_{1}$ and $p_{2}$ modes and their strange mode continuations at high temperatures. For both branches the horizontal area deformation is important in the  $3.6\lesssim \LTEF\lesssim 4.3$ range, and $\CHIE$ attains large positive values at high temperatures signifying substantial dominance of the compression term.}
	\label{fig:Fig3}
\end{figure} 

In \mFig{Fig3} the dependence of $\CHIE$ on $\LTEF$ is shown for the two overtone modes - $p_{1}$ and $p_{2}$ and for their continuation into the strange mode pair. Below $\LTEF\simeq 3.6$, $\CHIE$ is close to unity, indicating that the net contribution of the compression and horizontal area deformation terms is small relative to the inertial term. As the modes approach their transition into a strange-mode pair, their $\CHIE$ becomes positive. Along their strange-mode continuations, $\CHIE$ attains values of order $10$ and, for the excited branch, increases to nearly $30$ at the hot end of the sequence, showing that the compression term plays the dominant role in this range. In the intermediate temperature range $3.6\lesssim\LTEF\lesssim 4.3$ the $N_{\rm ha}$ term becomes dominant and causes $\CHIE$ to become large and negative ($\sim -10$) for the $p_{1}$ mode, while for the $p_{2}$ mode the $\CHIE$ is negative but stays closer to zero, in one interval becoming positive. Thus the compression and horizontal area deformation terms are more balanced for the $p_{2}$ mode with the $N_{\rm ha}$ being more dominant.

For both the damped and excited strange modes the $\CHIE>1$. So for the hot temperature strange modes the compression term $\NMC$ is larger in magnitude than the area deformation term $\NMH$. 

For the cold strange modes, which occur in the analyzed post-AGB envelope sequence for $\LTEF<3.96$ the $\CHIE$ is also positive, as may be seen from \mFig{Fig1}, except for the range $3.85\leq \LTEF \leq 3.91$ where $\CHIE<0$.

For the $S_{2}^+$ strange mode, for example, the compression and horizontal area distortion norms, as well as $\CHIE$ undergo a sudden and large change in this temperature range. Although the inertia norm also increases the change in the other two terms is orders of magnitude larger. The horizontal area deformation term becomes dominant and causes $\CHIE$ to attain very large negative values. This interval marks the onset of the strange mode transformation into ordinary p-mode. Near $\LTEF=3.9$ the mode is strange and its excitation rate attains maximum value, while near $\LTEF=3.95$ the mode already turns into damped ordinary mode. Thus this behavior may in part reflect the rapid rearrangement of eigenfunctions. 

At the same time, the surface contribution $\PDB$ for the mode increases by several orders of magnitude, and the ratio $|\PDB/\PEX|$ reaches approximately $0.25$. Together with the elevated spillover measures (Zalewski 2026b) in this range this suggests that the computed eigenmode becomes poorly conditioned and strongly sensitive to the boundary conditions. Ordinary modes in this temperature range do not exhibit an analogous disturbance in $\CHIE$. We therefore regard the negative values of $\CHIE$ as a manifestation of a broader boundary sensitive transition, rather than a property of cold strange modes.

\subsection{A criterion for negative $\CHIE$}
For radial pulsations, $N_{\rm sg}=0$, and the norm part of the balance relation may be written as
\[
\CHIE\, N_{\rm in} = N_{\rm in}+N_{\rm c}+N_{\rm ha}.
\]
Since $\NMI<0$, $\NMC<0$, and $\NMH>0$, we introduce the positive ratios
\[
\mathcal{R}_{j} = -\frac{N_j}{\NMH}, \qquad j\in \{\rm {in, c}\}.
\]
Then from the norm relation it is obtained that
\[
-\mathcal{R}_{\rm in}-\mathcal{R}_{\rm c}+1=-\CHIE\mathcal{R}_{\rm in},
\]
from which it follows that
\[
\CHIE=\frac{\mathcal{R}_{\rm in}+\mathcal{R}_{\rm c}-1}{\mathcal{R}_{\rm in}}.
\]
Since $\mathcal{R}_{\rm in}>0$, the condition for negative effective norm factor is
\[
\mathcal{R}_{\rm in}+\mathcal{R}_{\rm c}<1.
\]

Using the radial expressions for the norm terms these two ratios may be written as
\[
\begin{aligned}
	\mathcal{R}_{\rm in}&=\frac{|\sigma|^2}{4}\frac{\int C A_3 A_2 |d|^2\,dx}{\int C A_3 |d|^2\,dx} \\
	\mathcal{R}_{\rm c}&=\frac{1}{4}\frac{\int C A_4 |p|^2\, dx}{\int C A_3 |d|^2\, dx}.
\end{aligned}
\]

The first ratio contains explicit factor $|\sigma|^2$, and thus a mode with a small eigenfrequency modulus will tend to decrease its value. However, this ratio also depends on the spatial distribution of the displacement amplitude and on the structural factors $C$, $A_2$ and $A_3$ (see Zalewski 2026c). In low mass post-AGB envelopes $A_2\sim r^3$. The factor $A_2$ weights the numerator more strongly towards the surface. Consequently for a displacement perturbation that is less concentrated towards the outer layers the contribution weighted by $A_2$ decreases and thus the ratio may become smaller still.

The second ratio measures the compression contribution relative to the horizontal area deformation contribution. It depends on the amplitudes of displacement and pressure perturbations as well as on two factors $A_4$ and $A_3$. In the outer parts of the envelope, where both pressure and displacement perturbations may be large, $A_3>A_4$. Consequently the $\mathcal{R}_{\rm c}$ may remain smaller than unity unless pressure perturbation becomes sufficiently large relative to the displacement perturbation.

\subsection{Thermodynamic driving of a damped mode}
\begin{figure}[htb]
	\includegraphics{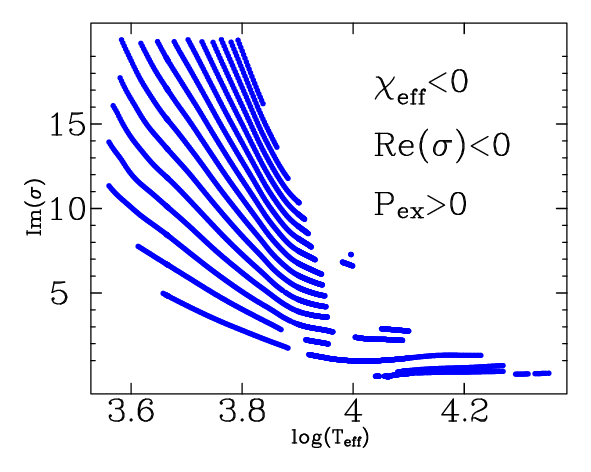}
	\FigCap{Oscillation frequency loci of damped modes with $\CHIE<0$ in the $\IMS$--$\LTEF$ plane. Only modes satisfying $|\PDB/\PEX|<0.05$ are shown. For these modes $\QEX>0$ follows from the balance relation, and since boundary contribution is small it follows that $\PEX>0$ as well. The overtone mode sequences belong to this class over most of the range $\LTEF<4$.}
	\label{fig:Fig4}
\end{figure}

In this section we shall analyze modes for which both $\CHIE<0$ and $\Re(\sigma)<0$. Modes of this kind may be found predominantly among ordinary mode overtones at lower effective temperatures ($\LTEF<4$). At higher temperatures individual lower frequency mode sequences can be identified as may be seen in \mFig{Fig4}.

From the balance relation 
\[
\QEX=2 \gamma \CHIE \EKIN
\]
it follows that when $\CHIE<0$ and $\RES<0$ then $\QEX>0$, since $\EKIN>0$. When the contribution from surface terms ($\PDB$) is small compared to $\PEX$ (as for the modes in \mFig{Fig4}) then 
\[
\QEX>0 \implies \PEX>0.
\]

Such modes are shown in \mFig{Fig4} - they are damped ($\RES<0$) although their thermodynamic exchange power ($\PEX>0$) makes a positive, driving contribution. 

From the total balance equation for these modes it follows that
\[
\PEX+\PRSP+\PDB=2\gamma \EKIN<0,
\]
indicates that the response contribution is negative and exceeds that of the exchange term. For radial damped modes $\mathcal{P}_{\rm c}>0$ and $\mathcal{P}_{\rm ha}<0$, hence it is the horizontal area deformation contribution that makes the $\PRSP$ negative. Thus while the conventional thermodynamic work would identify these modes as driven, its sign in fact does not determine the sign of the growth rate for these modes for the case when $\CHIE<0$.

\subsection{Thermodynamic damping of an excited mode}
\begin{figure}[htb]
	\includegraphics{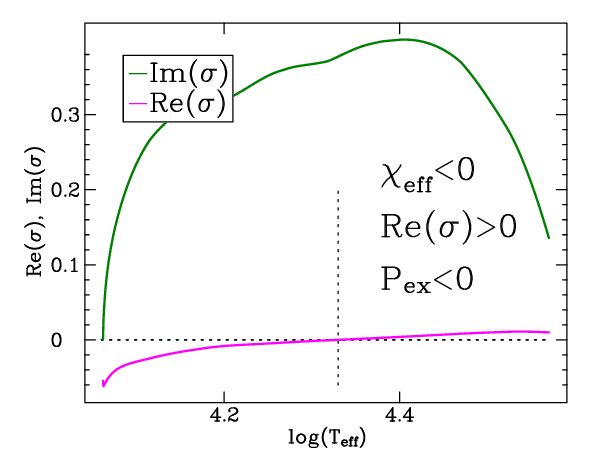}
	\FigCap{The oscillation frequency ($\IMS$) and excitation rate ($\RES$) of the mode as a function of envelope effective temperature. $\CHIE<0$ along this mode sequence. For $\LTEF>4.33$ (marked with dashed vertical line) the mode becomes excited.}
	\label{fig:Fig5}
\end{figure}

Since for $\CHIE<0$ there are numerous modes which have $\RES<0$ it is interesting to see whether there are modes for which $\CHIE<0$ and $\RES>0$. In the survey reported here we have found two such clear cases. Both are modes of low frequency and occur at temperatures $\LTEF>4$. The first case is a mode of frequency $\IMS\sim 1$ (at $\LTEF=4.4$). This mode becomes excited at temperatures $\LTEF\geq 4.48$. The second case is a mode with frequency $\IMS\sim 0.4$ (which at $\LTEF=4.4$ corresponds to a period of $\approx 2$ days) and it becomes excited for $\LTEF\geq 4.33$. The dependence of  $\sigma(\LTEF)$ for this mode is shown in \mFig{Fig5}.

As may be seen from \mFig{Fig5} the mode originates as damped thermal mode near $\LTEF=4.06$ and becomes oscillatory and then excited near $\LTEF=4.33$ after which it continues as excited mode. During the whole mode sequence $\CHIE<0$. 

\begin{figure}[htb]
	\includegraphics{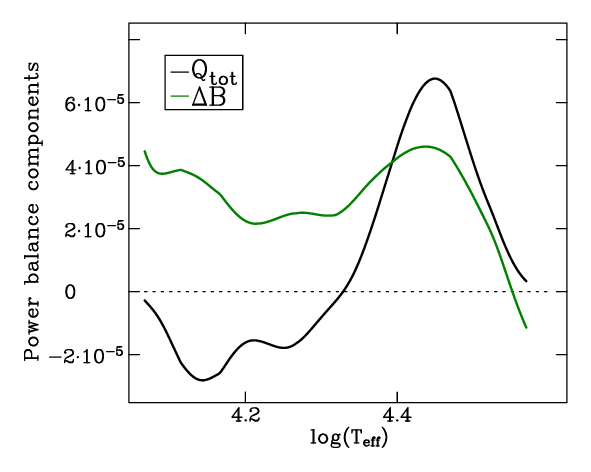}
	\FigCap{Comparison of the mode total power $\QTOT$ and surface terms $\PDB$ as a function of effective temperature. $\QTOT>0$ in the region where the mode is excited.}
	\label{fig:Fig6}
\end{figure}

The total power for the mode $\QTOT$ is given by
\[
\QTOT=\QEX+\PRSP.
\]
In \mFig{Fig6} the total power is shown for this mode sequence together with the surface term $\PDB$. For this mode, and other similar ones found in our survey the total power in the region where $\RES>0$ is small, of the order of the surface terms. For the analyzed mode it exceeds the surface terms for much of the hotter interval for $\LTEF\ge 4.4$. However, while the total power for the mode is small, it is positive, $\QTOT=2\gamma \EKIN>0$ since the mode is excited.

\begin{figure}[htb]
	\includegraphics{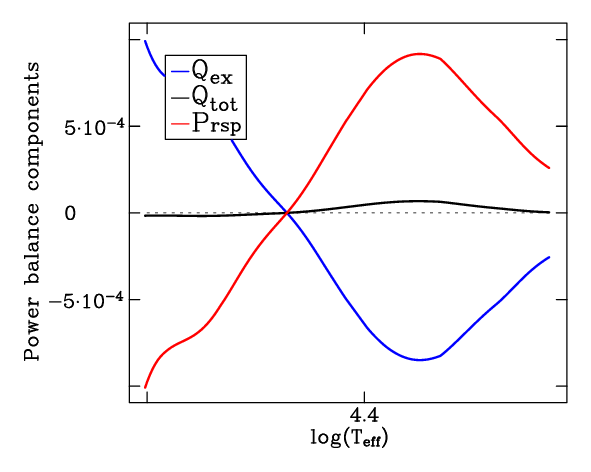}
	\FigCap{The components $\QEX$ - the exchange power including surface terms and $\PRSP$ - the response power of the total power $\QTOT$ as functions of effective temperature. When the mode becomes excited ($\QTOT>0$), the $\QEX$ becomes negative, while the response power becomes positive.}
	\label{fig:Fig7}
\end{figure}

Thus the analyzed mode is excited above $\LTEF\simeq 4.33$ and its total power $\QTOT>0$. In the region where the mode is excited the surface terms ($\PDB$) are small, of the order of $\QTOT$. A useful insight into the behavior of components of $\QTOT$ may be obtained using a plot of $\QEX$, $\PRSP$ and of the total power. This is presented in \mFig{Fig7}.

Since, as it is seen from \mFig{Fig7} the 
\[
\PRSP \gg \QTOT,
\]
in the region where $\CHIE<0$ and $\RES>0$. Then
\[
\QEX=\QTOT-\PRSP < 0.
\]

Therefore for the analyzed mode in the region where $\RES>0$, and away from $\RES=0$ where $|\QEX|$ becomes small, we have
\[
\QEX<0,
\]
and since, as may be seen from \mFig{Fig6}, $|\PDB|\sim |\QTOT|$ then
\[
|\PDB|\ll |\QEX|.
\]
Thus 
\[
\PEX<0.
\]

The same conclusion may be reached directly from 
\[
\QEX=2\gamma \CHIE \EKIN,
\]
which for the present case ($\CHIE<0$, $\gamma>0$) leads to $\QEX<0$ and thus to negative exchange power in this region.

Hence if the surface terms ($\PDB$) are small compared to the response and exchange terms then for regions where $\CHIE<0$ and $\RES>0$ the exchange term $\PEX<0$ and thus the thermodynamic term is indicating the mode is damped while in fact the mode is excited.

Therefore for excited modes for which $\CHIE<0$ and the surface terms are not dominating the $\PRSP$ and $\PEX$ the sign of the thermodynamic term does not determine the sign of the excitation rate.

\subsection{Dependence of modes with $\CHIE<0$ and $\RES>0$ on outer boundary conditions selector}

The calculations reported in the previous section were repeated using envelope model with same parameters but finer mesh - the number of layers used to compute pulsation for the low frequency mode for which $\RES>0$ while $\PEX<0$ and $\CHIE<0$ was increased from 23k to 107k layers using parameters $\{\alpha_{\rm c},\alpha_{\rm v},\alpha_{\rm k}\}=\{\texttt{2E-4},\texttt{5E-4},\texttt{0.5}\}$ (see Zalewski 2026b). The relative differences between the computed values, including $\QTOT$, are smaller than $10^{-4}$. Thus this suggests that the results for this mode are not caused by insufficient mesh resolution.

In Zalewski (2026a) it was shown that the spectra of radial modes in post-AGB envelope change substantially with the change of the form of outer boundary conditions. Therefore we have performed calculations for $\LTEF=4.4075$ using default mesh parameters, but changing the outer boundary conditions selector from $(3,4)$ used here to alternatives: $(1,2)$, $(2,3)$ and $(2,4)$. The first case corresponds to the outer boundary conditions determined from two fast (large) eigenvalues and their eigenvectors of the local dispersion relation at the surface, the second and third alternative forms include one fast branch of the local dispersion relation corresponding to the decaying or propagating outwards eigenvalue and one of the two with smallest modulus (see Zalewski 2026a for discussion). The form $(1,2)$ leads to large surface terms, as it admits an incoming fast branch, while the remaining two have smaller surface terms, comparable to the outer boundary selector made from the two slowest branches of the local dispersion relation used here $(3,4)$. The inner boundary selector in all cases was kept at $(1,3)$ representing a decaying or propagating inwards fast branch, and a slow branch.

\begin{figure}[htb]
	\includegraphics{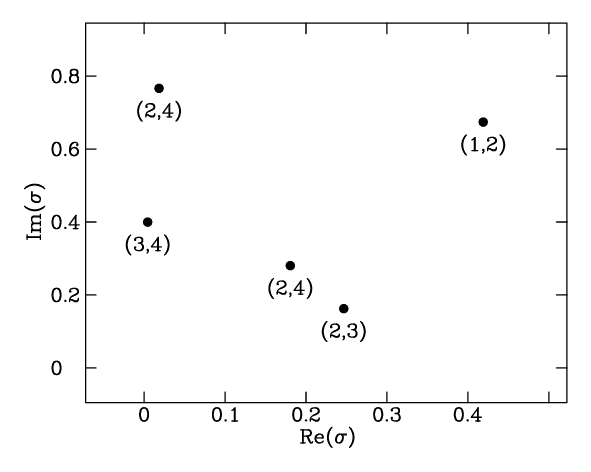}
	\FigCap{Excited low frequency radial modes with $\CHIE<0$. As may be seen the outer boundary conditions affect the position of the modes on the ($\RES$,$\IMS$)-plane, but excited modes exist for all examined types of outer boundary conditions.}
	\label{fig:Fig8}
\end{figure}

In \mFig{Fig8} the unstable low frequency modes for $\LTEF=4.4075$ satisfying $\CHIE<0$ and obtained for these four types of outer boundary selectors are shown.

For $(2,4)$ there are in fact two excited low frequency modes, while for $(1,2)$ the excitation rate of the mode is large. Since however for the $(1,2)$ selector the outer surface terms are substantial the mode found is not considered a reliable candidate for low frequency excited mode example. From this figure it is seen that the spectrum of low frequency modes is sensitive to the selected form of outer boundary conditions.

Although the detailed spectrum depends strongly on the adopted outer boundary conditions, excited modes with $\CHIE<0$ persist for all boundary formulations examined. For the \((2,4)\) and $(2,3)$ selectors, similarly as for $(3,4)$, both surface terms are small. Therefore the reversed relation between the thermodynamic exchange power and growth rate persists under changes in outer boundary conditions.

\section{Conclusions}

We have examined the quadratic balance relation introduced for radial nonadiabatic pulsations in Zalewski (2026c) and generalized to nonradial pulsations in Zalewski (2026d). The relation has been applied to radial modes of a sequence of post-AGB envelope models spanning $3.52\leq\LTEF\leq4.6$.

The total power governing the growth or damping of a mode, including the surface contribution, may be written as
\[
\QTOT = \PEX + \PRSP + \PDB = 2\gamma\EKIN,
\]

where $\gamma=\RES$. The exchange power, defined as the thermodynamic exchange power together with the surface contribution, satisfies
\[
\QEX = \PEX + \PDB = 2\gamma\CHIE\EKIN.
\]

The effective norm entering this relation is
\[
\NEF=\NMI+\NMC+\NMH+\NMS, \qquad \CHIE=\frac{\NEF}{\NMI},
\]
where $\NMS=0$ for radial pulsations.

For nearly adiabatic modes, $\CHIE$ is expected to be close to unity. However in strongly nonadiabatic pulsations the compression ($\NMC$) and horizontal area deformation ($\NMH$) contributions need not be small compared with the inertia term ($\NMI$). In particular, $\NMH$ can dominate the combined inertia and compression terms, causing $\NEF$ and $\NMI$ to have opposite signs and hence resulting in $\CHIE<0$.

Since $\EKIN>0$, the sign of $\QEX$ is determined by the product $\gamma\CHIE$. Consequently, when $\CHIE<0$, the usual relation between the sign of the thermodynamic exchange power and that of the growth rate is reversed, provided that the surface contribution is sufficiently small.

In the cooler models, with $\LTEF<4$, a broad family of radial overtone modes (\mFig{Fig4}) has $\CHIE<0$ and $\gamma<0$. For these modes $\QEX>0$ and, when $\PDB$ is negligible, $\PEX>0$. The thermodynamic exchange therefore drives these modes, although they remain damped because the response power provides a larger opposing contribution (\mFig{Fig2}). The negative effective norm factor is caused by the dominance of the horizontal area deformation term.

Away from the narrow, boundary sensitive transition near $\LTEF\simeq 3.9$, the strange modes considered here generally have positive values of $\CHIE$, which may substantially exceed unity when the compression contribution dominates (\mFig{Fig3}). The sudden large and sign changing values found during the transformation of a cold strange mode into an ordinary p-mode coincide with enhanced surface terms and spillover measures. They are therefore regarded as a property of this boundary-sensitive transition rather than as a general property of cold strange modes.

In the hotter models, with $\LTEF>4$, we have identified two low frequency modes for which $\CHIE<0$ and $\gamma>0$. In this case $\QEX<0$ and, away from $\gamma=0$ where the surface contribution becomes relatively important, $\PEX<0$. Thus the thermodynamic exchange damps these modes, whereas the positive response power is sufficiently large to make the total power positive and the modes are excited.

The results for the low frequency mode were reproduced using a substantially finer mesh. The frequencies and number of excited modes depend strongly on the adopted outer boundary conditions, but excited solutions with $\CHIE<0$ were found for every boundary formulation examined. For the outer boundary selectors of the $(2,3)$, $(2,4)$, and $(3,4)$ types, the explicit surface terms remain small. The reversed relation between thermodynamic exchange power and growth rate therefore cannot be attributed solely to an explicit boundary contribution.

These results show that, in strongly nonadiabatic pulsations, the sign of the thermodynamic exchange power alone does not determine whether a mode is excited or damped. The response terms and, in the radial problem particularly, the term associated with horizontal area deformation, can reverse the conventional relation between thermodynamic driving and the mode growth rate. The low frequency excited modes found here may consequently provide an additional route to linear pulsational instability outside the ordinary p-mode instability region. Whether these modes persist in complete stellar models and under closer examination of the atmospheric boundary treatment remains to be established.

\end{document}